# The Evaluation of a New Daylighting System's Energy Performance: Reversible Daylighting System (RDS)

Masoome Haghani[1], Behrouz MohammadKari[2], and Rima Fayaz[3]
[1]NCState University, Raleigh, NC
[2]Building and Housing research Center (BHRC), Tehran, Iran
[3]Art University of Tehran, Tehran, Iran

## ABSTRACT

This paper evaluates the energy performance of a new daylighting system, patented by the author, in a regular closed office space. The advantage of this new system as opposed to conventional venetian blinds is its rotating capability, which improves the energy efficiency of the space. Computer simulation method has been conducted to examine the performance of this new system on the south aperture of a closed-office space with 30% Window to Wall ratio (WWR) in three cities in Iran with different climate zones based on ASHRAE: Tehran (3B), Tabriz (4B), and Yazd (2B). The simulation has been implemented in Honeybee platform with EnergyPlus engine to simulate the combined total load consisting of heating, cooling, and lighting loads. To control lighting, a dimming control is applied to the space. The results of the study represent the benefits of the reversible daylighting system (RDS) over the state-of-the-art venetian blinds to improve the energy efficiency of the space through just changing the location of the blind during heating/cooling demand time of the year.

**Keywords**  Window Detail, Daylighting system, Energy Efficiency, Thermal Load, Office, Climate zone

## INTRODUCTION

Passive energy-saving strategies in buildings help to improve the global warming effect of fossil fuels. The thermal transfer between indoor and outdoor mostly depends on windows when the opaque facade is well insulated. Among passive design strategies, the use of shading devices is crucial in sustainable building design to control solar heat gains and daylighting conditions. Venetian blinds are the common daylighting systems to control penetrating solar radiation especially on the south façade of buildings. Different studies have been conducted to improve the efficiency of Venetian blinds by studying their properties or/and optimizing blind slats controls: Tzempelikos studied the geometry and tilt angle of venetian blinds on view and light (Tzempelikos 2008). Oh et al. evaluated the double sided blind and the effect of automated control on energy performance (Oh, Lee, and Yoon 2012). Naderi et al. investigated the optimization of controlled blind specification to reduce the energy consumption, and thermal and visual discomfort (Naderi et al. 2020). Also, several studies have worked on the blind control system through machine learning and ANN based methods (Luo et al. 2021; Yeon et al. 2019).

Meanwhile, few studies have evaluated the location of blinds and its effect on energy efficiency of buildings.

For instance, Yoon et al. studied the effect of blind reflectance on heating and cooling load regarding location (either inside or outside) and glazing type (Yoon, Kim, and Lee 2014). They recommend low SHGC windows integrated with the low reflectance exterior blinds in the case of cooling dominant buildings such as office buildings and those under hot climatic conditions and low U-value windows integrated with the high reflectance interior blinds in the case of heating dominant buildings such as residential buildings and those under cold climatic conditions. Haghani et al. evaluated the energy efficiency of horizontal and vertical blinds in different orientations and location of blinds (Haghani, Kari, and Fayaz 2017) and recommended the best slat angle of blinds for heating and cooling time periods in office buildings.

The advantage of this new daylighting system (RDS) as opposed to conventional venetian blinds is the location change capability which improves the energy efficiency of the system. The system is consisted of a UPVC casement frame with compression seal technology to seal the window from air leakage potential, and a UPVC pivoted window outfitted with a blind which has 360 degrees rotation freedom to be located inside or outside of the window based on the energy demand of the space (exterior blind in cooling demand time and interior blind in heating demand period). The rotation of the pivoted system is easily applicable and users could physically rotate it like a pivoted window without too much effort. Figure 1 represents a schematic view of the system and figure 2 is the mock-up model of the system.

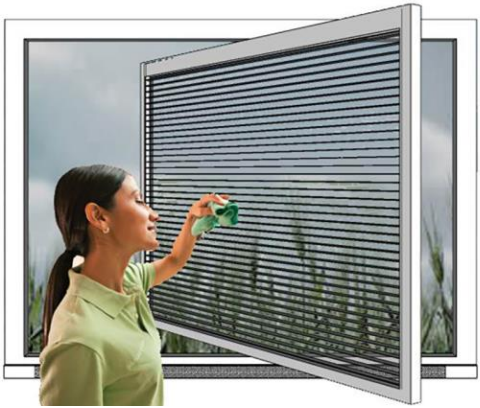

*Figure 1: Schematic figure of the new daylighting system*

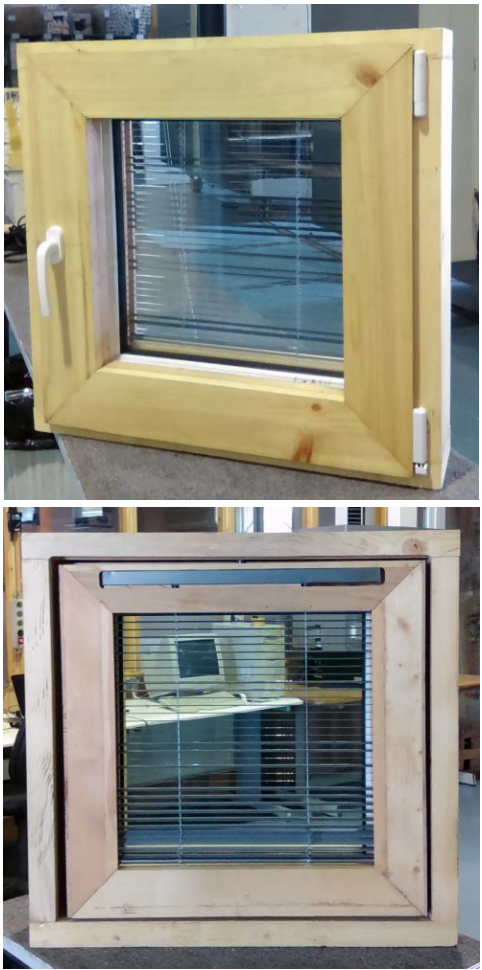

*Figure 2: Mock-up model of the new daylighting system; Upper image: Casement frame, Bottom image: Pivoted window outfitted with blind*

Computer simulation method has been conducted to examine the performance of this new system on the south aperture of a closed-office space with 30% Window to Wall ratio (WWR) in Tehran, Iran (climate zone 3A, based on ASHRAE standard (ASHRAE90.1 2019)). The simulation has been implemented in Honeybee platform with EnergyPlus engine to simulate the combined total load consisting of heating, cooling, and lighting loads. To control lighting, a dimming control is applied to the space.

In this study we focus on the performance of this new daylighting system to investigate its energy performance in reducing the total load of office buildings (heating, cooling, and lighting load). The proposed RDS system will be evaluated and compared with two regular daylighting systems: Interior horizontal blind, and exterior horizontal blind.

## METHODOLOGY

To evaluate the energy performance of the new daylighting system and comparing it with conventional horizontal blinds, a computer simulation study conducted for three cities in Iran with different climates. Table 1 shows the climatic zones of each city based on ASHRAE 90.1 classification. According to ASHRAE 90.1, Tehran is located in climate zone 3B which is defined as warm-dry climate, Yazd is located in climate zone 2B which is defined as hot-dry climate, and Tabriz is located in climate zone, 4B defined as mixed-dry climate (Briggs, Lucas, and Taylor 2003).

Since commercial buildings have a big window-to-wall ratio and a common application of Venetian blind is in this type of buildings, an office room has been selected for this study and modeled in Rhino 6.0. The simulation has been conducted in Honeybee platform by OpenStudio software, which utilizes EnergyPlus engine. The analysis has been performed for different locations of blinds on south facade with recommended blind slat angle in the author's previous study: 30° in cooling demand time and 120° in heating demand period (Haghani, Kari, and Fayaz 2017). Figure 3 shows the different slat angles of a blind.

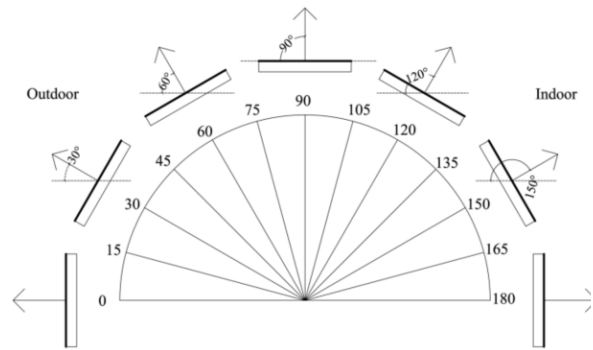

*Figure 3: Slat different angles*

Table 1: Selected cities climate zone

|  | Tehran | Tabriz | Yazd |
|---|---|---|---|
| Climatic Zone | 3B | 4B | 2B |
| Latitude | 35.42 | 38.13 | 31.90 |
| Longitude | 51.15 | 46.23 | 54.28 |

## Room Description

A typical office room of dimensions 3.5 m (width)×4:0m (depth)×3:0m (height) has been considered in the model. It is a south oriented room with WWR 30%; a schematic overview of simulation model in Rhino environment is illustrated in figure 4.

Table 2: Exterior skin thermal properties

| City | | Btu/sfK | Value |
|---|---|---|---|
| Tehran | Exterior Wall | U-value | 0.077 |
| | Window (WWR=30%) | U-value | 0.42 |
| | | SHGC | 0.25 |
| Yazd | Exterior Wall | U-value | 0.084 |
| | Window (WWR=30%) | U-value | 0.45 |
| | | SHGC | 0.25 |
| Tabriz | Exterior Wall | U-value | 0.064 |
| | Window (WWR=30%) | U-value | 0.36 |
| | | SHGC | 0.36 |

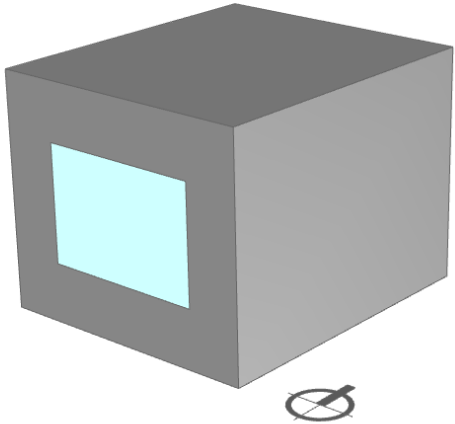

Figure 4: 3D model of the simulated room

This space is designed for two people based on the publication 178 of Plan and Budget Organization of Iran: Office Building Design Regulation (of Housing and Design 1998). The floor, ceiling, and walls, except for the south wall, have been considered adiabatic to not have any heat transfer with the outdoors. Table 2 demonstrates the U-value of exterior skin in different climate zones based on ASHRAE 90.1-2019 (ASHRAE90.1 2019).

Table 3 represents internal gain and temperature setpoints for heating and cooling of 21◦C and 26◦C, respectively, based on the General Design Criteria of ASHRAE Handbook HVAC Applications (2016). The thermal load profiles have been calculated using the EnergyPlus function "Zone-HVAC: IdealLoadsAir System" without modeling the heating and cooling systems. This object provides the required supply air capacity to each zone at user specified temperature and humidity ratio to calculate the heating and cooling loads (Yoon, Kim, and Lee 2014).

## Blind Specification

Two locations have been considered for blinds with the same specifications represented in figure 5: interior blind and exterior blind. Table 5 demonstrates the specification of blind for this study.

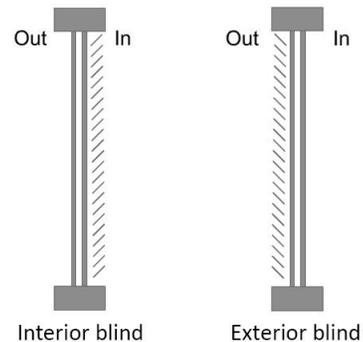

Figure 5: Blind location

## RESULTS AND DISCUSSION

### Heating, cooling, and lighting load with blind location and reflectance variation

The annual heating, cooling, and lighting load of the space has been evaluated considering 30◦ and 120◦ for slats angle during cooling and heating period, respectively, with different blind locations (interior and exterior blind), and different slat reflectance (0.1, 0.5, and 0.9). The result is illustrated in figure 6. In the case of interior blind, the annual heating load in all climate zones and slat angles increases as the slat reflectance enhances. On the other hand, the cooling load in all climate zones and slat angles decreases as the slat reflectance increases. Finally, the lighting load in all climate zones and slat angles decrease by increasing the slat reflectance.

Table 3: Input conditions of internal heat gain and indoor set-points

| | Input Data | | Reference |
|---|---|---|---|
| Internal Gain | People | 2 person | |
| | Lighting | 7.97 (W/m2) | ASHRAE 90.1 (2019) |
| | Equipment | 6.89 (W/m2) | EnergyPlus default |
| Temperature Set-point | Cooling | 26°C | ASHRAE Handbook – HVAC Applications (2007) |
| | Heating | 21°C | |

Table 4: Selected cities climate data

| | Dry Bulb Temp | | | | | | | | | | |
|---|---|---|---|---|---|---|---|---|---|---|---|
| Tehran | 3.89 | 4.98 | 10.66 | 16.3 | 21.74 | 28.26 | 29.87 | 30.07 | 25.9 | 17.79 | 11.41 | 5.59 |
| Yazd | 7.11 | 9.98 | 16 | 21 | 25.77 | 31.4 | 33.17 | 31.28 | 27.63 | 21.03 | 12.86 | 7.06 |
| Tabriz | -1.52 | 0.52 | 6.55 | 11.78 | 17.45 | 22.96 | 26.27 | 26.9 | 22.13 | 14.96 | 6.21 | 0.21 |
| | Humidity | | | | | | | | | | |
| Tehran | 60.04 | 56.82 | 49.55 | 38.91 | 34.1 | 21.92 | 25.67 | 24.48 | 25.75 | 38.24 | 49.15 | 63 |
| Yazd | 49.38 | 39.66 | 26.93 | 30.86 | 21.66 | 14.22 | 15.62 | 13.07 | 15.82 | 21.69 | 33.16 | 46.45 |
| Tabriz | 64.17 | 66.05 | 56.26 | 52.61 | 49.35 | 37.98 | 34.95 | 31.21 | 36.15 | 47.84 | 63.1 | 66.41 |

Table 5: Simulated blind properties

| Field | | Value |
|---|---|---|
| Width (mm) | | 20 |
| Distance (mm) | | 20 |
| Conductivity (W/mK) | | 160 |
| Reflectance | Front | 0.5 |
| | Back | 0.5 |
| Emissivity | Front | 0.9 |
| | Back | 0.9 |
| Slat Angle (Fixed) | Heating Period (°) | 120 |
| | Cooling period (°) | 30 |

In the case of exterior blind, the annual heating load in all climate zones and slat angles decreases with an increase in the slat reflectance. On the other hand, the cooling load in all climate zones and slat angles increases by the slat reflectance increase. Finally, the lighting load in all climate zones and slat angles decrease as the slat reflectance enhances.

To sum up, increasing the slat reflectance results in a decreasing pattern in cooling and lighting load, and an increasing pattern in heating load in the case of interior blind. In the case of exterior blind, this pattern inverses; an increase in the slat reflectance brings about heating and lighting load decrease and cooling load increase. In other words, heating and cooling load behaves inversely in the case of interior and exterior blind. Meanwhile, regardless of blind location, lighting load decreases when the reflectance of slat increases. These results are in accordance with the Yoon et al. study (Yoon, Kim, and Lee 2014).

Table 6 demonstrates the total load of space in different cities under different blind locations, and slat reflectance. As mentioned before, in all cases, 30° and 120° have been considered for slat angle during the cooling and heating demand period, respectively. Therefore, in the case of interior blind, the high reflective slats have a better performance (1447.39 kWh, 1253.11 kWh, and 1531.4 kWh respectively for Tehran, Tabriz and Yazd) and in the case of exterior blind, it is inverse. (1315.05 kWh, 1135.36 kWh, and 1402.05 kWh respectively for Tehran, Tabriz and Yazd). Figure 7 represents the breakdown of annual load (kWh) for different cities in the case of interior blind with slat reflectance 0.9 and the exterior blind with slat reflectance 0.1.

**Window heat gain analysis**

To better understand the background behind the pattern of heating and cooling of the space, window heat gain analysis has been conducted with EnergyPlus software to calculate the transferred solar heat through the window with shading devices. The solar-optical model for blinds in EnergyPlus is based on (Lomanowski 2009;

*Table 6: The total load of space in different cities under different blind locations, and slat reflectance*

| City | Slat Reflectance | Interior Blind | Exterior Blind |
|---|---|---|---|
| Tehran | 0.1 | 1720.72 | 1315.05 |
|  | 0.5 | 1615.24 | 1361.83 |
|  | 0.9 | 1447.39 | 1384.48 |
| Tabriz | 0.1 | 1461.38 | 1135.36 |
|  | 0.5 | 1382.82 | 1213.01 |
|  | 0.9 | 1253.11 | 1221.36 |
| Yazd | 0.1 | 1770.01 | 1402.05 |
|  | 0.5 | 1673.14 | 1446.41 |
|  | 0.9 | 1531.40 | 1471.51 |

H.Simmler, U.Fischer, and F.Winkelmann 1996; DOE 2018). The total heat gain through window outfitted with blind depends on the blind specification (blind location, material, slat angle); therefore, it significantly affects the thermal load of the space. The total heat gain of the window includes different heat transfer based on the location of the blind. The equation (1) and equation (2) represent the total heat gain of the window respectively for the interior blind and exterior blind:

$$Q_{window} = Q_{solartrans} + Q_{conv.air} + Q_{conv.blind} + Q_{rad.win} + Q_{rad.blind} Q_{rad.out} + Q_{cond.frame} \quad (1$$

$$Q_{window} = Q_{solartrans} + Q_{conv.win} + Q_{rad.win} - Q_{rad.out} + Q_{cond.frame} \quad (2$$

Where, $Q_{window}$ is the total window heat gain, $Q_{solartrans}$ is the solar radiation transmittance through the window, $Q_{conv.air}$ is the convective heat flow to the zone from the air flowing through the gap between glazing and interior blind, $Q_{conv.blind}$ is the convective heat flow to the zone from the interior blind, $Q_{rad.win}$ is the net infrared heat flow to the zone from the window, $Q_{rad.blind}$ is the net infrared heat flow to the zone from the interior blind, $Q_{rad.out}$ is the shortwave radiation from zone transmitted back out the window, $Q_{cond.}$ frame is the conduction to zone from window frame and divider and $Q_{conv.win}$ is the convective heat flow to the zone from the window (Yoon, Kim, and Lee 2014).

Figure 8 represents the total heat gain of the window in different cities and under different conditions. Regarding the location of the blind, the total heat gain for the interior blind is bigger in value than the exterior blind. This value is about two times higher for the interior blind than the exterior blind when the slat angle is 30◦. Meanwhile for the blind with slat angle 120◦, this difference is smaller in amount since the slats are mostly open. On the other hand, considering the slat reflectance, the pattern of window heat gain decreases by increasing the slat reflectance in the case of interior blind and the inverse pattern occurs in the case of exterior blind. This result is in accordance with (Yoon, Kim, and Lee 2014; Yoon, Yun, and Kim 2015).

The pattern of heat gain of the window in figure 8 justifies the heating and cooling load of the space in figure 6; to provide thermal comfort for the space, the greater heat gain the window system receive results in the lower heating load for heating demand period and more cooling load for the cooling demand period. This result confirms the importance of the effect of window heat gain on heating and cooling load of the space.

**Comparison of RDS system performance with the conventional Venetian blinds**

The result of annual thermal and lighting load of space with the Reversible Daylighting System (RDS) has been taken from the same previous simulations because the only difference between the RDS system and conventional blinds is the capability to change the location of blind, inside or outside of the window, depending on the space heating/cooling demand. In the case of RDS system, the space has an interior blind for heating demand period (from November to April) and an exterior blind for cooling demand period (from May to October). Therefore, comparing figure 7 and figure 9, the heating load and lighting load of the space is lower than the space outfitted with exterior blind for the entire year and the cooling load of the space is less than the space outfitted with interior blind for the entire year. In other words, the RDS system is more energy efficient than the interior blind during cooling demand period of the year, and is more energy efficient than the exterior blind during the heating demand period of the year. Since in the case of heating load, the low reflective blind has the lowest load and in the case of exterior blind, the low reflective blind has the minimum cooling load, the low reflective slats have the best performance in reducing the total load of the office space (Figure6) for the RDS system. Figure 9 demonstrates the heating, cooling, and lighting load of the space in different climate zones when the low reflective RDS system has been installed on the window. During heating demand of the space, the blind is located inside the space and the annual heating load is equivalent to the interior blind case with slat reflectance

0.1, 72.6 kWh, 210.1 kWh, and 36.24 kWh for Tehran, Tabriz, and Yazd respectively. During cooling demand of the space, the blind is located outside and the annual cooling load is equivalent to the exterior blind case with slat reflectance 0.1; 964.76 kWh, 642.27 kWh, and 1082.01 kWh for Tehran, Tabriz, and Yazd respectively.

In the next step, the total load of the office space for the RDS system has been compared with interior and exterior blinds in different climate zones which is shown in figure 10. As the results represent, in all cities the RDS system has better performance in reducing the total load of the space. In addition, the efficiency of the RDS system is evaluated based on the conventional blinds and the result is demonstrated in figure 11.

Figure 11 shows that the RDS system performs 9% and 3% more efficient comparing the interior and exterior blind respectively in Tehran with warm-dry climate (Climate zone 3B). In Tabriz with mixed-dry climate (climate zone 4B), its performance is 9% and 6% better than interior and exterior blind respectively. In Yazd with hot-dry climate (climate zone 2B), the RDS system performs 8% and 2% better than interior and exterior blind respectively.

Based on this comparison, RDS system performs better than both interior and exterior blinds and this improvement is greater in the case of interior blind than the exterior blind. In addition, the RDS system has a better performance in mixed-dry climate like Tabriz than the hot-dry climate like Yazd as shown in figure 11.

## CONCLUSION

This study evaluated the new daylighting system, Reversible Daylighting System (RDS). To this end, in the first step, the heating, cooling, and lighting load difference was investigated under blind space and slat reflectance variation in slat angle 120° during heating load and 30° during cooling period. Three different ASHRAE climate zones were considered for this study; Tehran (climate zone 3B), Tabriz (climate zone 4B), and Yazd (climate zone 2B). The sensitivity analysis was conducted to evaluate the effect of different reflectance blind slats. The result was in accordance with the previous study which has been done in this area.

In the second step, the space load was evaluated based on the blind location, and the slat reflectance. The result represented that in the case of interior blind the high reflective blind performs better and in the case of exterior blind the low reflective blind performs better.

Next, the load of the office space was investigated for the RDS system based on the two previous steps' results. The results show the low reflective blind performs the best in the case of the RDS system. Finally, the performance of the RDS system was compared with conventional blinds. The results showed that in the RDS system, there was more efficiency improvement in comparison with the interior blinds than the exterior blinds. Furthermore, this system performs better in a mixed-dry climate zone than the hot-dry climate zone.

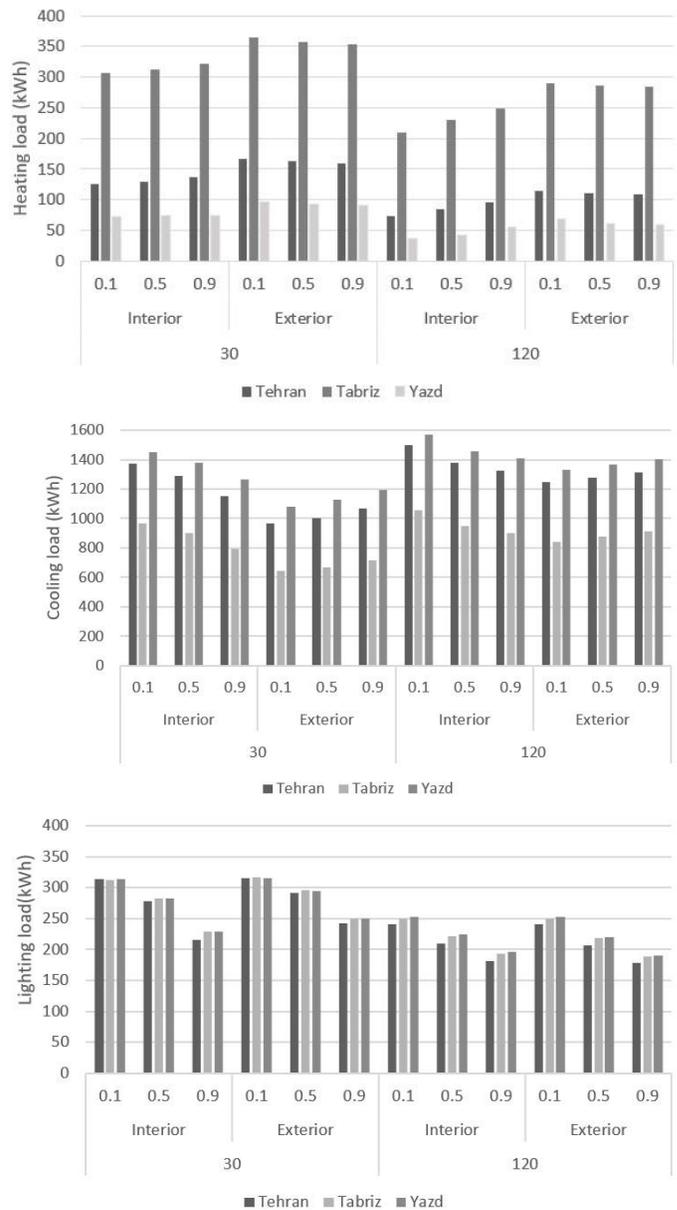

*Figure 6: Blind location and slat reflectance impact on annual heating, cooling, and lighting load*

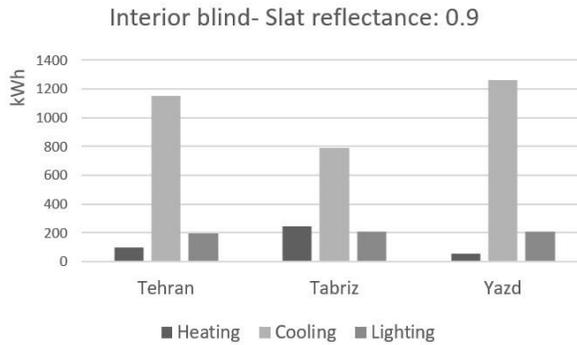
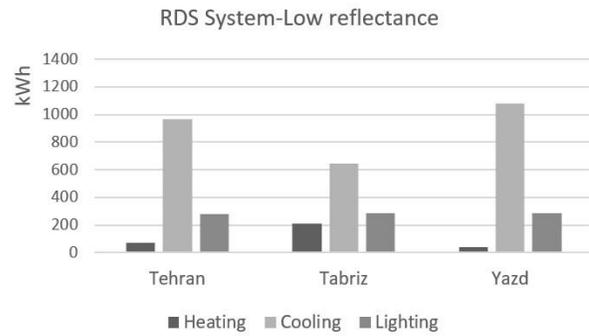

*Figure 9: Annual heating, cooling, and lighting load in different cities*

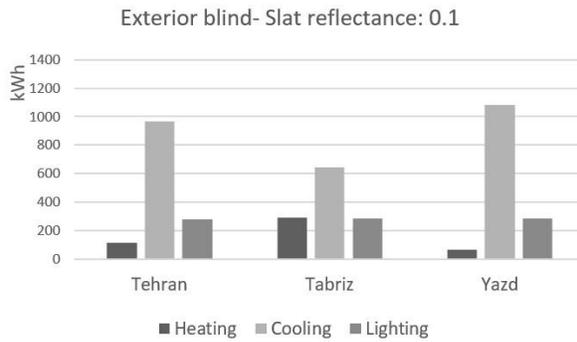

*Figure 7: Annual heating, cooling, and lighting load in different cities*

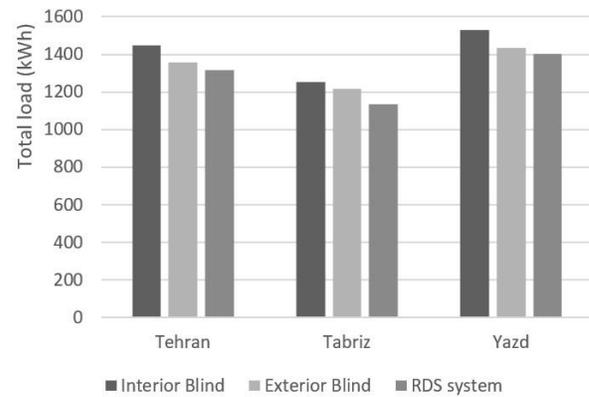

*Figure 10: Comparing the total load in different cities*

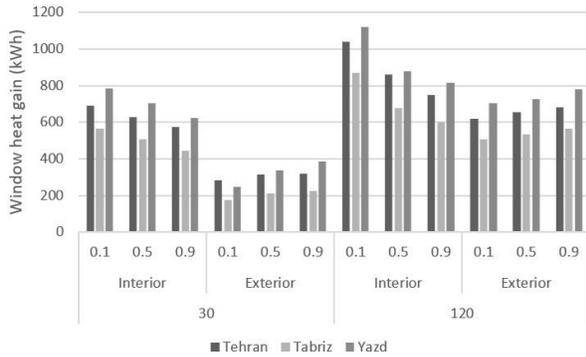

*Figure 8: Window heat gain variation considering different blind reflectance, slat angle, and location in different cities*

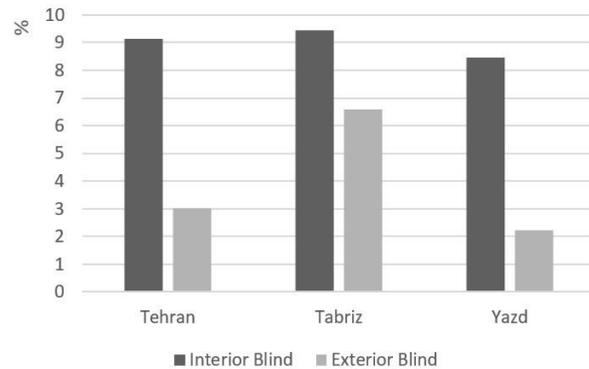

*Figure 11: Comparing the efficiency of RDS system with interior and exterior blind in different cities*

**Limitations**

The main limitation of this study is evaluating the RDS system in only office buildings and the research could be expanded to residential and educational buildings. This study focused on simulations and verifying the results based on the previous studies, since different building types have completely different thermal behavior. In addition, the impact of automatic slat angle control could also be studied, since automatic slat control might lead to better performance of the RDS system. Another potential study could investigate the cost efficiency of the RDS system in comparison with the state of the art blind shadings. Additional climate zone assessment would also be beneficial to cover all eight climate zones of ASHRAE classification in other countries, as this study just evaluated three climate zones in Iran.